\documentclass[preprint,aps,showpacs,superscriptaddress,tightenlines,a4paper]{revtex4}

\usepackage{amssymb}
\usepackage{amsmath}
\usepackage{bbold}
\usepackage{mathrsfs}
\usepackage{paralist}
\usepackage{graphicx}
\usepackage{xcolor}

\newcommand{\ii}{{\rm i}}
\newcommand{\e}{{\rm e}}

\begin{document}
%
%
    \title{Wave propagation in complex coordinates}
%
%
	\author{S. A. R. Horsley}
	\affiliation{Department of Physics and Astronomy, University of Exeter,
Stocker Road, Exeter, EX4 4QL}
	\email{s.horsley@exeter.ac.uk}
	\author{C. G. King}
	\affiliation{Department of Physics and Astronomy, University of Exeter,
Stocker Road, Exeter, EX4 4QL}
	\author{T. G. Philbin}
	\affiliation{Department of Physics and Astronomy, University of Exeter,
Stocker Road, Exeter, EX4 4QL}
%
%
    \begin{abstract}
    We investigate the analytic continuation of wave equations into the complex position plane.  For the particular case of electromagnetic waves we provide a physical meaning for such an analytic continuation in terms of a family of closely related inhomogeneous media.  For bounded permittivity profiles we find the phenomenon of reflection can be related to branch cuts in the wave that originate from poles of \(\epsilon(z)\) at complex positions.  Demanding that these branch cuts disappear, we derive a large family of inhomogeneous media that are reflectionless for a single angle of incidence.  Extending this property to all angles of incidence leads us to a generalized form of the P\"oschl Teller potentials.  We conclude by analyzing our findings within the phase integral (WKB) method. 
    \end{abstract}
%
%
    \maketitle
%
%
	The propagation of linear waves through inhomogeneous media is straightforward to compute numerically, but there are comparatively few useful analytical solutions (in 1D probably restricted to the solutions of Riemann's differential equation~\cite{dlmf2015}).  The ease of numerical simulation is in some respects unfortunate because it leads to the belief that everything is well understood about linear wave equations.  Finding solutions may be numerically trivial, but it is almost impossible to make any general statements on this basis.  In this respect transformation optics~\cite{pendry2006,leonhardt2006a} has been a big step forward, providing a simple intuitive recipe for relating inhomogeneous materials to analytical solutions (revealing hidden symmetries of apparently unsymmetrical structures~\cite{kraft2014}).  It is worth keeping in mind that concealment devices only became a concrete possibility~\cite{pendry2006,leonhardt2006b,leonhardt2008} on the basis of this mathematical insight that coordinate deformations are equivalent to inhomogeneous media.
	\par
	Aspects of transformation optics have been known for some time: Tamm explored the connection between wave propagation in anisotropic media and curved geometries in the 1920s~\cite{tamm1924}.  The perfectly matched layer~\cite{berenger1994} is another familiar example, and it was established some time ago that a complex transformation of the spatial coordinates is equivalent to an inhomogeneous non--reflecting absorbing medium~\cite{chew1997}.  More recently there has been a revived interest in the idea of using complex coordinates to understand wave propagation~\cite{popa2011,castaldi2013,orlov2014}, as well as the discovery of a relationship between media with loss and gain and invisibility~\cite{longhi2011,lin2011,mostafazadeh2013}, and an improvement in the practical realisation of materials with tailored dispersion and dissipation~\cite{ye2013}.  Our work is similar in spirit, as we also make use of complex coordinates to understand the effect of inhomogeneous material properties on the propagation of waves.  However, rather than use complex coordinate transformations to derive different sets of material properties, we seek to understand the behaviour of the wave in the entire complex position plane, applying this as a method for understanding wave propagation in a family of inhomogeneous media.
%
%
	\begin{figure}[ch!]
		\includegraphics[width=9cm]{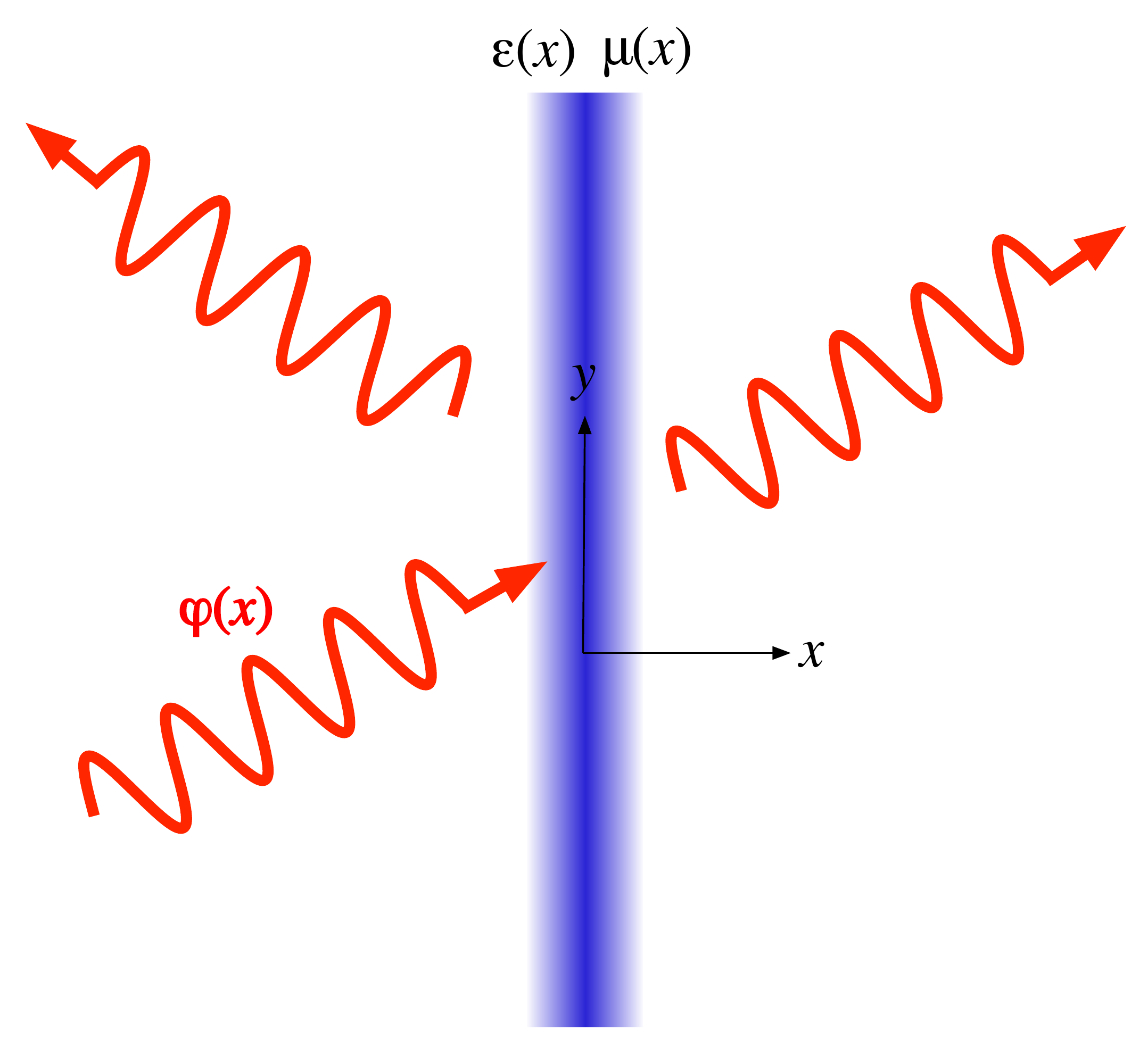}
		\caption{The solution to the monochromatic wave equation (\ref{helmholtz}) represents a TE polarized wave propagating at an angle (determined by \(k_{y}\)) through an inhomogeneous medium defined in terms of the permittivity and permeability, \(\epsilon(x)\) and \(\mu(x)\).  In this work we examine the analytic continuation of this wave propagation into the complex \(x\) plane, and through doing this find profiles \(\epsilon(x)\) (\(\mu=1\)) that do not reflect radiation, in certain cases independent of the angle of incidence.\label{schematic_figure}}
	\end{figure}
	\par
	The key message is that it can be useful to analytically continue a wave equation into complex coordinates, and that we can for instance find families of reflectionless inhomogeneous media through doing this.  The particular wave equation we shall use to demonstrate this is the equation for the electric field of a monochromatic (frequency \(\omega\)) TE polarized wave \(\hat{\boldsymbol{z}}\cdot\boldsymbol{E}=\varphi\) propagating through an isotropic inhomogeneous slab of material
	\begin{equation}
		\left[\frac{d}{dx}\left(\mu^{-1}(x)\frac{d}{dx}\right)+k_{0}^{2}\epsilon(x)-k_{y}^{2}\mu^{-1}(x)\right]\varphi(x)=0\label{helmholtz}
	\end{equation}
	where propagation occurs in the \(x\)--\(y\) plane, \(k_{0}=\omega/c\), \(k_{y}\) determines the angle of incidence, and the permeability and permittivity depend on \(x\) (see figure~\ref{schematic_figure}).  Equation (\ref{helmholtz}) also holds for the TM polarization, with the roles of \(\epsilon\) and \(\mu\) interchanged.  Although we have written (\ref{helmholtz}) as an equation for the electromagnetic field, the same equation occurs in other areas of wave physics.  For example an acoustic wave obeys (\ref{helmholtz}), but the bulk modulus and the density play the role of the permittivity and permeability.
%
%
	\section{A physical meaning for complex coordinates}
	\par
	Although (\ref{helmholtz}) holds for a wave propagating along the real axis \(x\in(-\infty,\infty)\), we can analytically continue \(\varphi\) into the complex position plane \(z=x_{1}+{\rm i}x_{2}\).  The immediate thought might be that this is nothing more than mathematical pontification.  Yet there is a well defined physical meaning for the whole complex position plane, which we now describe.  
	\par
\begin{figure}[ch!]
		\includegraphics[width=8.1cm]{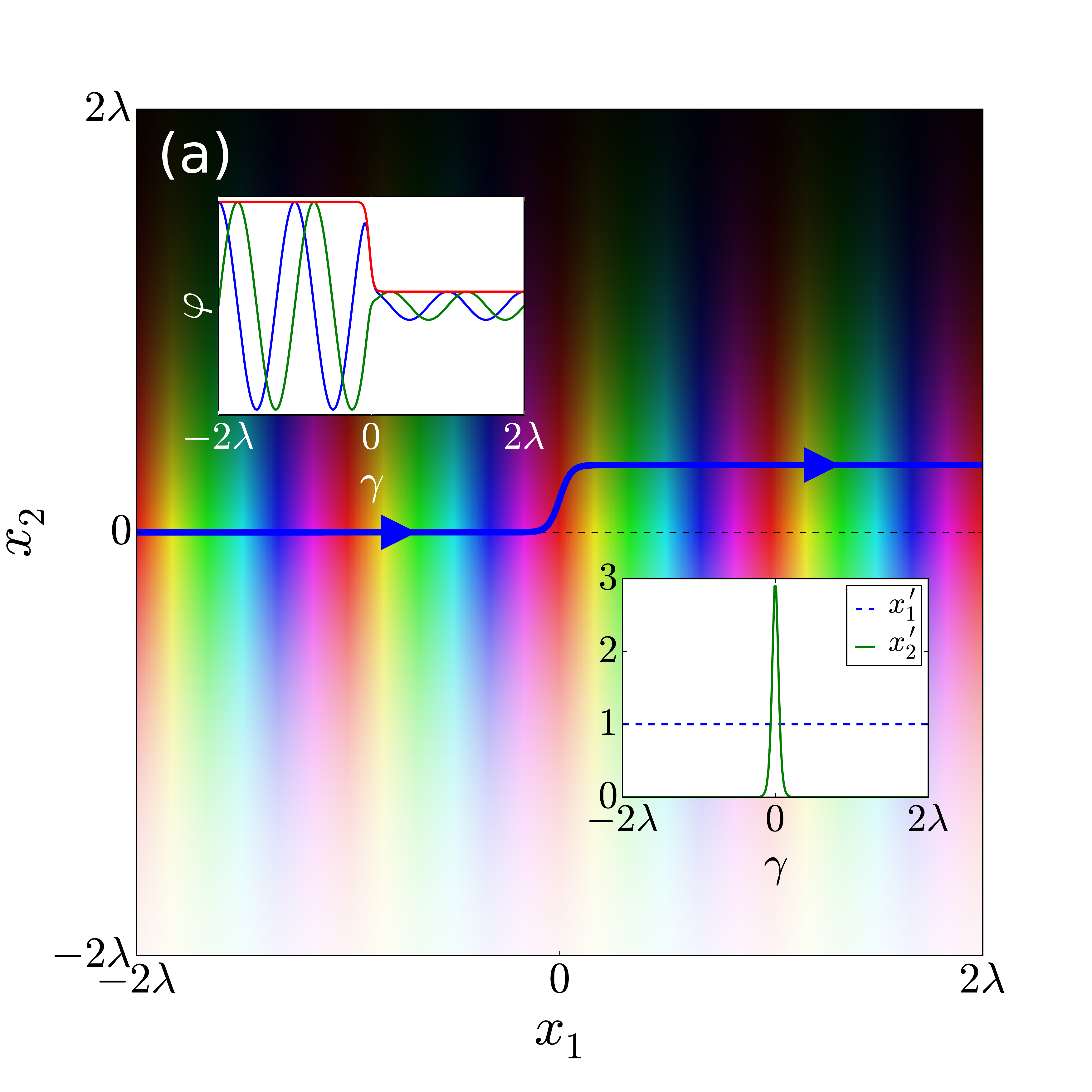}
		\includegraphics[width=8.1cm]{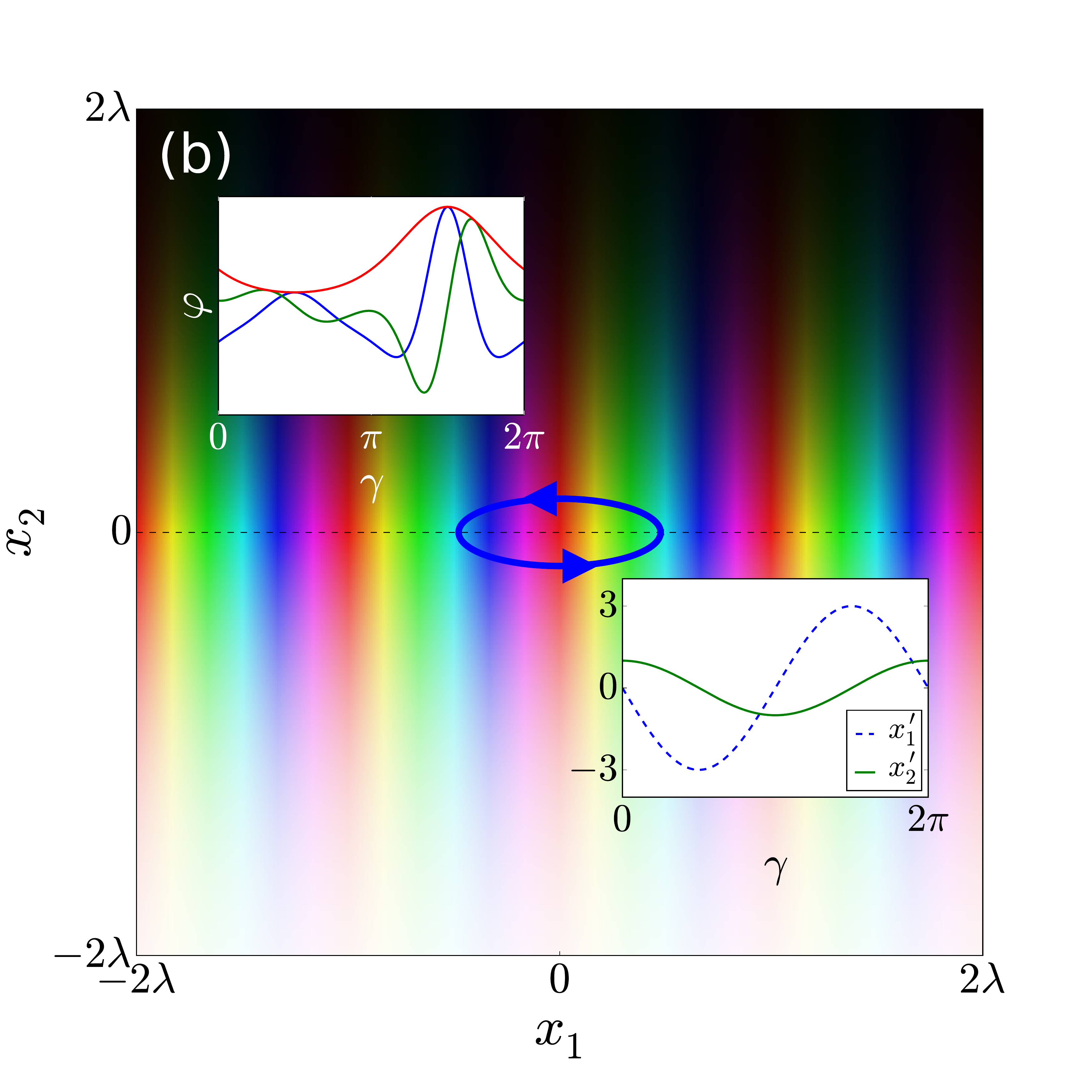}
		\caption{Two examples showing the equivalence between wave propagation along trajectories in complex coordinates and in inhomogeneous media. In both panels we plot the wave \(\varphi(z)=\exp(2\pi {\rm i} z/\lambda)\) as a function of \(z=x_{1}+{\rm i}x_{2}\) with brightness indicating magnitude and colour indicating phase: \([0,\pi/2,\pi,3\pi/2]\) represented as [red,green,cyan,purple].  The top subfigure within each panel shows \(\varphi\) evaluated along the contour \(z(\gamma)\) shown in blue (in each top subfigure blue is the real part, green the imaginary part, and red the magnitude), and the bottom subfigure shows \(z'(\gamma)\), which gives the equivalent material parameters via (\ref{materials}) with \(\epsilon(\gamma)=\mu(\gamma)=1\). (a) The contour \(z(\gamma)=\gamma+{\rm i}\left(1+\tanh(3\gamma)\right)\) represents an inhomogeneous reflectionless absorber; (b) the closed contour \(z(\gamma)=3\cos(\gamma)+{\rm i}\sin(\gamma)\) represents a parity--time symmetric periodic medium where the wave propagates with a phase that advances then reverses through the unit cell.\label{complex_path_figure}.}
	\end{figure}

	Writing (\ref{helmholtz}) along a trajectory in the complex plane \(z=z(\gamma)\) parametrized by a real number \(\gamma\) the equation becomes
	\begin{equation}
		\left[\frac{d}{d\gamma}(z')^{-1}\mu^{-1}(\gamma)\frac{d}{d\gamma}+k_{0}^{2}z'\epsilon(\gamma)-k_{y}^{2}\mu^{-1}(\gamma)z'\right]\varphi(\gamma)=0\label{complex-path}
	\end{equation}
	where \(z'=dz/d\gamma\).  If we interpret \(\gamma\) as a new position variable, this is equivalent to propagation through a new inhomogeneous anisotropic medium where
	\begin{align}
		\epsilon_{zz}(\gamma)&=\epsilon(\gamma)z'(\gamma)\nonumber\\
		\mu_{xx}(\gamma)&=\mu(\gamma)(z'(\gamma))^{-1}\nonumber\\
		\mu_{yy}(\gamma)&=\mu(\gamma)z'(\gamma).\label{materials}
	 \end{align}
	 In general these material parameters will be complex functions of position, and the medium will exhibit some combination of dissipation and gain.  The equivalence (\ref{materials}) has an interesting consequence: having solved the wave equation as a function of \(x\), analytic continuation into the complex position plane is equivalent to solving the wave equation in an infinite number of closely related inhomogeneous media, one for every trajectory \(z(\gamma)\).  The material parameters will be generally anisotropic, unless \(z(\gamma)\) is parallel to the real axis.  Figure~\ref{complex_path_figure} shows the simplest case of such an analytic continuation, \(\varphi(x)\to\varphi(z)=\exp({\rm i}k z)\) with two examples to illustrate the equivalence of complex trajectories and materials.  Note that the equivalence is not restricted to this one dimensional example, and also holds for the three dimensional wave equation, as is well known by those working on the theory of perfectly matched layers~\cite{chew1997}.  The substitution \(\varphi\to\sqrt{\mu}\varphi\) reduces equation (\ref{helmholtz}) to the form \(\varphi''+(k_{0}^{2}\epsilon_{\text{eff}}-k_{y}^{2})\varphi=0\), where \(k_{0}^{2}\epsilon_{\text{eff}}=k_{0}^{2}\epsilon\mu-(3/4)(\mu'/\mu)^2+(1/2)\mu''/\mu\), and \(\mu'=d\mu/dz\).  For the remainder of this work we therefore consider---\emph{without loss of generality}---the case \(\mu=1\) in (\ref{helmholtz}), understanding \(\epsilon\) as \(\epsilon_{\text{eff}}\).
%
%
	\section{Poles in the permittivity and branch cuts in the wave\label{pole-section}}
	\par
	Having provided a physical motivation for the analytic continuation of \(\varphi\) into the complex position plane, we free ourselves to investigate the general behaviour of \(\varphi\) as a function of \(z\).  In particular we shall show how the phenomenon of reflection manifests itself in the complex position plane, and how it can be controlled through the function \(\epsilon(z)\).
	\par
	For the sake of simplicity we restrict this investigation to bounded permittivity profiles tending to unity at large \(|z|\), and we choose to represent such an \(\epsilon(z)\) as an infinite product
	\begin{equation}
		\epsilon(z)=\prod_{i=0}^{\infty}\frac{z-q_{i}}{z-p_{i}}\label{eps_gen}
	\end{equation}
	where the zeros \(q_{i}\) and poles \(p_{i}\) are all confined to a region \(|z|<|z_{\text{max}}|\).  The behaviour of the wave \(\varphi\) in the complex position plane has some generic features that arise from the positions and weights of the poles of (\ref{eps_gen}).
	\par
	We have found that the effect of the poles in \(\epsilon(z)\) on the wave \(\varphi(z)\) is to introduce branch cuts that run from the poles to infinity.  This can be simply understood within the Born approximation~\cite{wu2011}, which we apply along lines of constant \(x_{2}\) in the \(z\) plane (Green function: \(\exp(\ii k|x_{1}-x_{1}'|)/2\ii k\)).  As \(x_{1}\to\pm\infty\) a wave incident from the left can be approximated to
	\begin{equation}
		\varphi(x_{1}+{\rm i}x_{2})\sim{\rm e}^{{\rm i}k(x_{1}+{\rm i}x_{2})}+
		\begin{cases}
			\frac{{\rm i}k_{0}^{2}}{2k}{\rm e}^{-{\rm i}k x_{1}}{\rm e}^{-k x_{2}}\int_{-\infty}^{\infty}dx'[\epsilon(x'+{\rm i}x_{2})-1]{\rm e}^{2{\rm i}kx'}&x_{1}\to-\infty\\[10pt]
			\frac{{\rm i}k_{0}^{2}}{2k}{\rm e}^{{\rm i}k x_{1}}{\rm e}^{-k x_{2}}\int_{-\infty}^{\infty}dx'[\epsilon(x'+{\rm i}x_{2})-1]&x_{1}\to\infty
		\end{cases}	
		\label{born}
	\end{equation}
	with \(k=(k_{0}^{2}-k_{y}^{2})^{1/2}\).  Given that \(k>0\), the integration contour in the first of the expressions (\ref{born}) can be closed in the upper half \(x'\) plane, which gives a null result if \(\epsilon(z)\) is analytic in the region \({\rm Im}[z]>x_{2}\), and a non--zero result if \(\epsilon(z)\) is not analytic in this region.  Note that there is a subtlety here: if the permittivity approaches unity as \(1/z\) or slower then the waves won't be plane waves at infinity, as assumed in (\ref{born}) and in these cases (\ref{born}) neglects a logarithmic term in the phase.  Nevertheless the equation correctly reproduces the first order reflection and transmission coefficients (see  appendix~\ref{appA}).
	\par
\begin{figure}[ch!]
		\includegraphics[width=10cm]{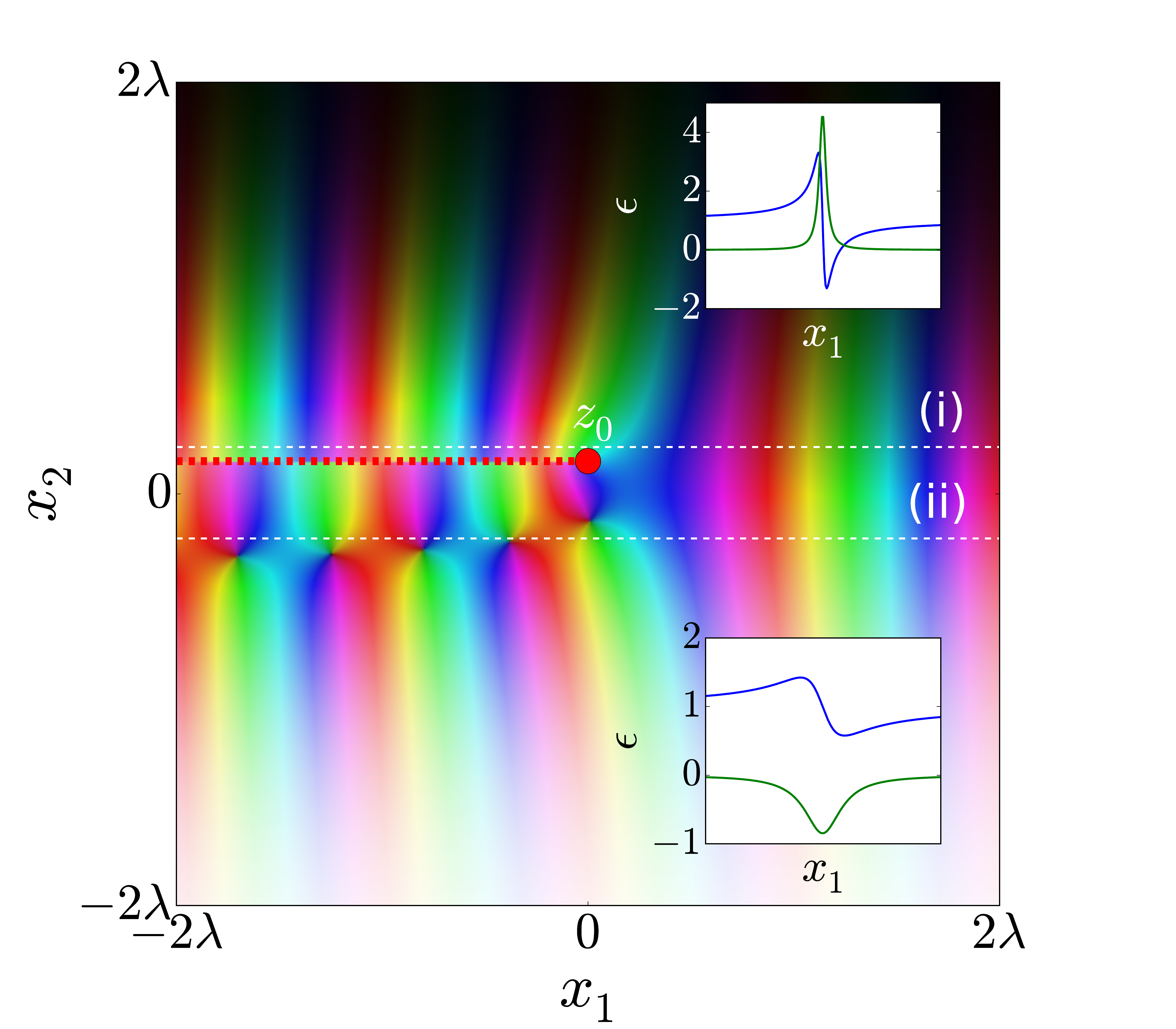}\\
		\includegraphics[width=6.5cm]{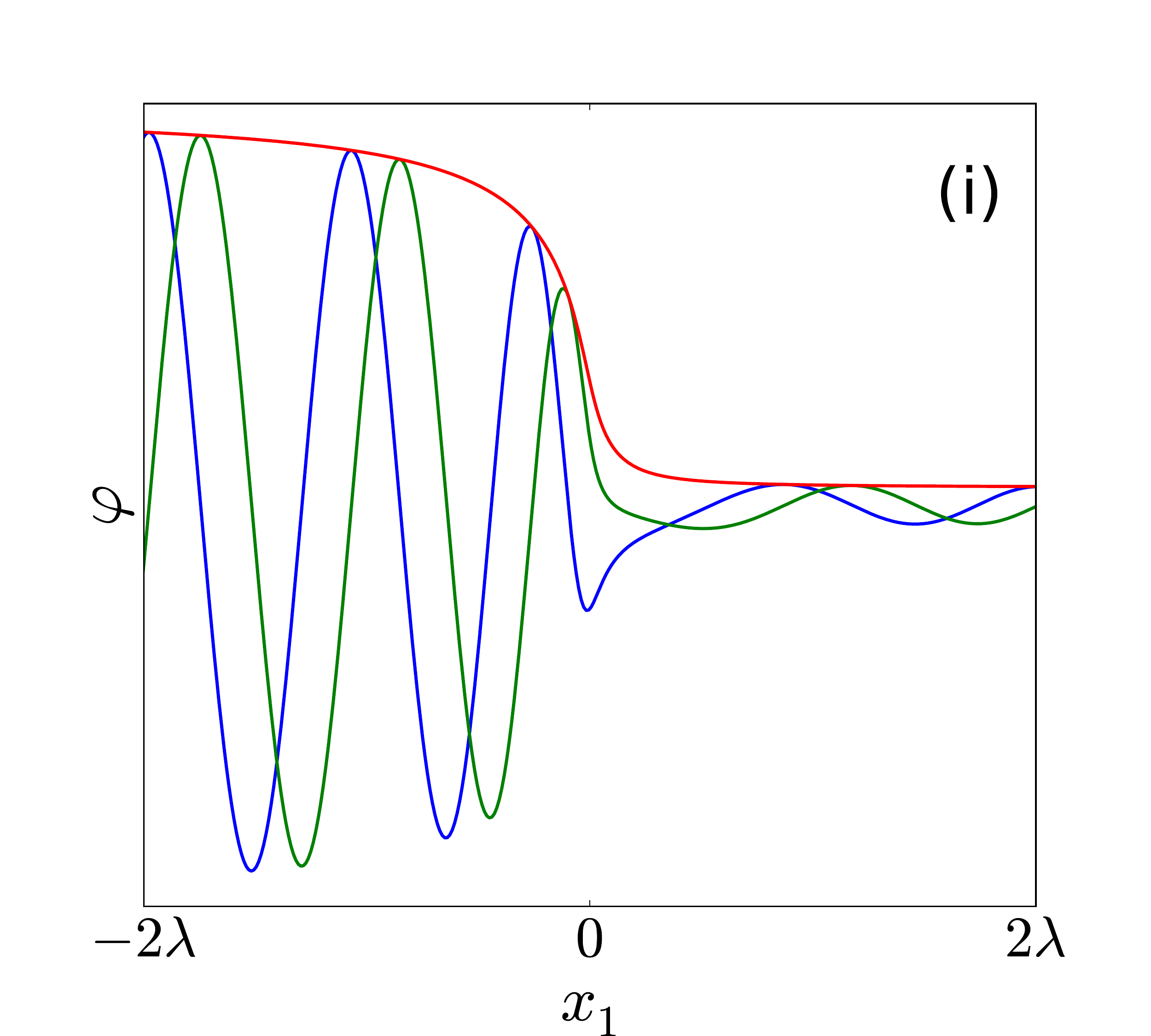}\includegraphics[width=6.5cm]{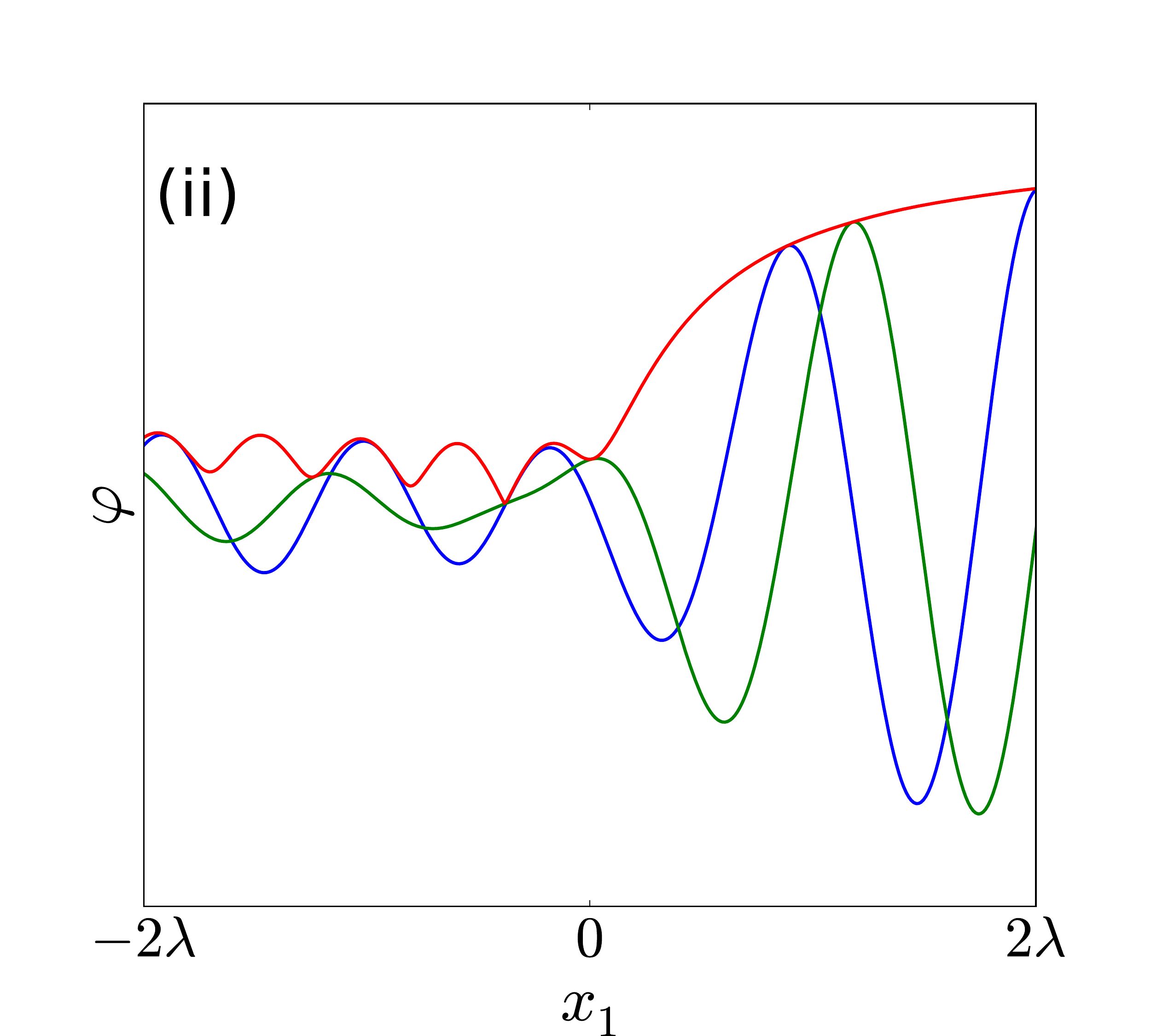}
		\caption{Numerical solution of (\ref{helmholtz}) for the function \(\varphi(z)\) in the profile \(\epsilon(z)=1-2a/(z-z_{0})\), \(\mu=1\), for the case \(k_{y}=0\), \(a=1\), \(k_{0}=1\), and \(z_{0}=\ii\).   The numerical solution was obtained through imposing the boundary condition \(\varphi(z)=1\) and \(\varphi'(z)={\rm i}\sqrt{\epsilon(z)}k_{0}\) at some position on the far right of the profile where \(\epsilon'(z)\sim0\), and then numerically integrating (\ref{helmholtz}) along a series of lines parallel to the real axis.  The red dot at \(z_{0}\) indicates the pole in the permittivity profile, and the red dashed line indicates a branch cut.  The two lower panels (i) and (ii) show the functional form of \(\varphi(z)\) evaluated along the white dashed lines shown in the upper panel (colour conventions as in figure~\ref{complex_path_figure}) and the inset panels show the spatial variation of \(\epsilon\) along these lines.  The wave in panel (i) is evaluated in a region just above the branch cut where the medium is dissipative, and despite the rapid variation of \(\epsilon(z)\) there is no reflection, evident in the lack of oscillations in the magnitude of the wave where \(x<0\).  Panel (ii) shows the wave evaluated below the branch cut in a region where the medium is amplifying: the oscillations in the absolute value of the field on the left show that here there is significant reflection.\label{branch_cut_fig}}
	\end{figure}

	
	In the simplest case of a single pole \(\epsilon(z)=(z-q_{0})/(z-p_{0})\) the first of (\ref{born}) evaluates to
	\begin{equation}
		\varphi(x_{1}\to-\infty)\sim{\rm e}^{{\rm i}kz}+{\rm e}^{-{\rm i}kz}
		\begin{cases}
			\frac{\pi k_{0}^{2}}{k}(q_{0}-p_{0}){\rm e}^{2{\rm i}kp_{0}}&{\rm Im}[p_{0}]>{\rm Im}[z]\\
			0&{\rm Im}[p_{0}]<{\rm Im}[z]
		\end{cases}\label{reflection-complex-plane}
	\end{equation}
	Meanwhile on the far right of the profile, \(x_{1}\to\infty\)
	\begin{equation}
		\varphi(x_{1}\to\infty)\sim{\rm e}^{{\rm i}kz}+\frac{\pi k_{0}^{2}}{2k}(q_{0}-p_{0}){\rm e}^{{\rm i}kz}
		\begin{cases}
		1&{\rm Im}[p_{0}]>{\rm Im}[z]\\
		-1&{\rm Im}[p_{0}]<{\rm Im}[z]\label{transmission-complex-plane}
		\end{cases}.
	\end{equation}
	Applying the boundary condition \(\varphi(x_{1}\to\infty)\sim\exp{({\rm i}kz)}\), it is clear that \(\varphi(z)\) has a branch cut that runs horizontally left from the pole at \(z_{0}\).  If the wave had been incident from the right, then the branch cut would run to the right.  For the remainder of this work we restrict ourselves to the case of incidence from the left, with the understanding that similar results hold for the other direction.  Figure~\ref{branch_cut_fig} shows an illustration of this phenomenon, where we have numerically integrated the wave equation and confirmed the presence of the branch cut for this simple case.  In appendix~\ref{appA} we give an analytic example, also confirming the existence of the branch cut and the validity of (\ref{reflection-complex-plane}--\ref{transmission-complex-plane}).
%
%
%
%
	\begin{figure}[hc!]
		\includegraphics[width=14cm]{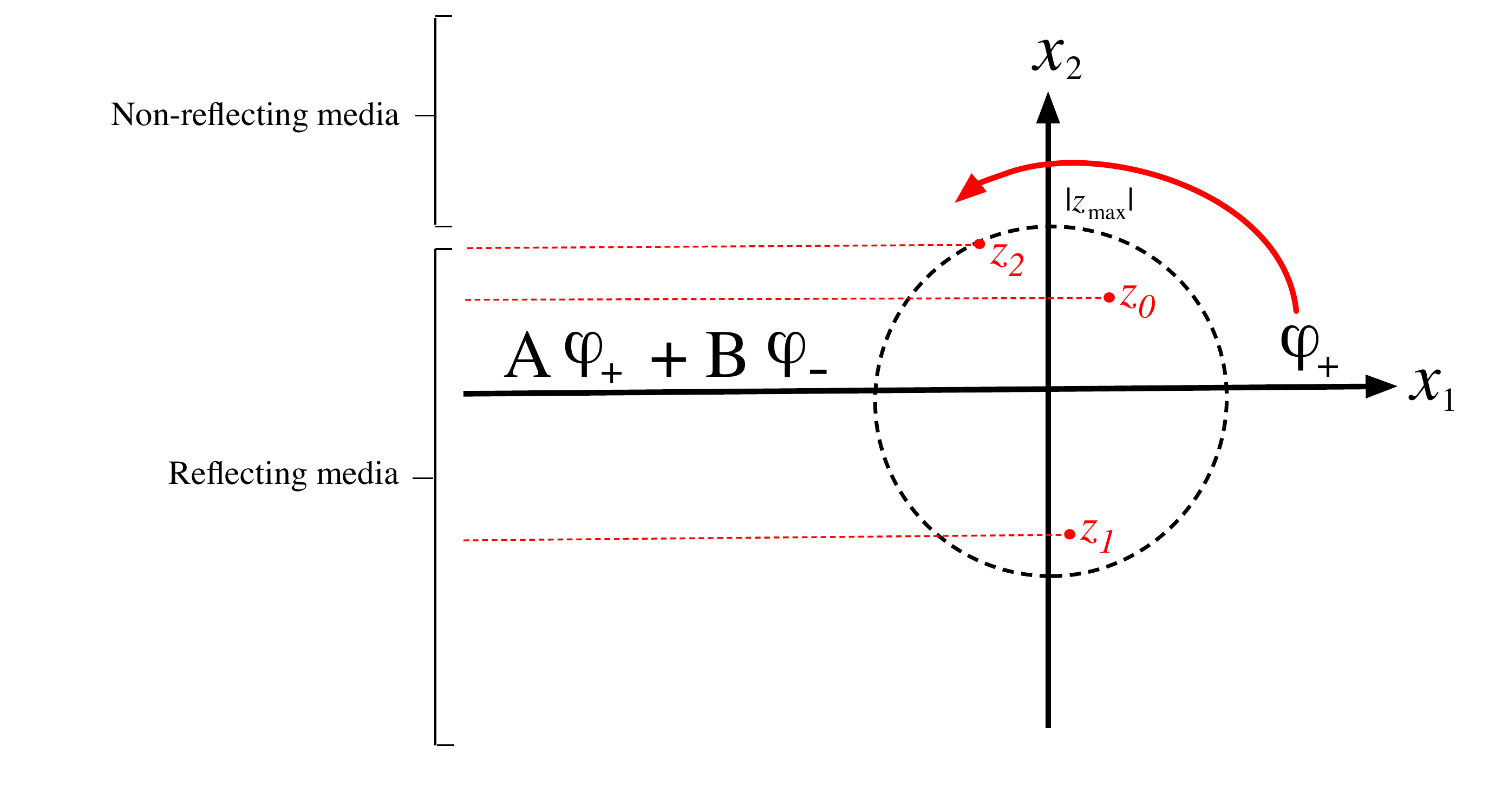}
		\caption{Imposing the boundary condition that the wave is right--going on the far right of the profile \(\varphi=\varphi_{+}\), and integrating the wave equation (\ref{complex-path}) around a large semicircle in the upper half plane \(z(\gamma)=R\exp{({\rm i}\gamma)}\), we find that the wave becomes exponentially small.  Meanwhile, were we to follow the same procedure on the left and follow a sum of left and right--going waves \(\varphi=A\varphi_{+}+B\varphi_{-}\) along \(z(\gamma)\) in the clockwise sense then we would conversely find that the solution was exponentially large in the upper half plane.  The presence of branch cuts (red dashed lines emerging from the poles of \(\epsilon(z)\)) in \(\varphi\) avoids this contradiction.  The wave is analytic in the region of the complex position plane \(x_{2}>|z_{\rm max}|\), where it also tends to zero as \(x_{2}\to\infty\).  All the inhomogeneous media captured in this region are reflectionless.\label{complex_plane_fig}}
	\end{figure}
	\par
	In general a branch cut in \(\varphi(z)\) runs from each of the poles of \(\epsilon(z)\) to infinity, and we shall now argue that the presence of the cuts is due to the phenomenon of reflection.  Firstly we note that in complex analysis, Liouville's theorem~\cite{cartan1995} shows that every bounded entire function is constant.  This means that an inhomogeneous permittivity \(\epsilon(z)\) that tends to a constant value at large \(|z|\) cannot be an entire function and must contain poles.  Now consider the illustration given in figure~\ref{complex_plane_fig}.  Suppose we move out to the semi--circle \(|z|\to\infty\) in the upper half plane, and consider wave propagation along the contour \(z(\gamma)=R\exp{({\rm i}\gamma)}\).  Using (\ref{materials}) we find that this is equivalent to propagation through a medium where \(\epsilon_{zz}=[1+{\rm O}(1/R)]{\rm i}R\exp{({\rm i}\gamma)}\), \(\mu_{yy}={\rm i}R\exp{({\rm i}\gamma)}\) \& \(\mu_{xx}=-{\rm i}\exp{(-{\rm i}\gamma)}/R\).  This is a material with a negative index and a large degree of gain in the region \(\gamma\in[\pi/2,\pi]\) and a negative index and a large degree of dissipation in the region \(\gamma\in[0,\pi/2]\).  Starting from the boundary condition that the wave is right--going at \(\gamma=0\), and integrating through the dissipative region, the wave is exponentially diminished at \(\gamma=\pi/2\).  Meanwhile, starting from the boundary condition of a sum of right and left going waves at \(x_{1}=-\infty\), and integrating through the gain medium, the wave is exponentially amplified at \(\gamma=\pi/2\), unless \(B=0\).  This leads to a contradiction unless there is a branch cut in \(\varphi\) that forces the reflected wave to disappear past a certain value of \(\gamma\).  This is the jump captured in expression (\ref{reflection-complex-plane}).  Note that the branch cut can be moved to run from the pole to infinity in any way we choose, although it is non--physical to consider wave propagation along a contour that passes across the cut.  Moreover this shows that above all of the cuts the reflection disappears, even though the medium may be rapidly varying in this region of the complex position plane.
	\par
	 We can use this understanding to find a family of inhomogeneous media that reflect no radiation for any angle of incidence---a finding already reported in~\cite{horsley2015}, but derived in a different way.  If we consider the region of complex position above all of the poles in \(\epsilon(z)\), \(x_{2}>|z_{\rm max}|\) (see figure~\ref{complex_plane_fig}) and the branch cuts in \(\varphi\) run parallel to the \(x_{1}\) axis, then in that region \(\varphi(z)\) is an analytic function that tends to zero as \(x_{2}\to\infty\).  In analogy to what is typically done in the frequency domain~\cite{volume5}, such a function can be represented as a Fourier integral over positive wave--numbers
	\[
		\varphi(z)=\int_{0}^{\infty}\frac{dk}{2\pi}\tilde{\varphi}(k){\rm e}^{{\rm i}k(z-z_{\rm max})}.
	\]
	The lack of any negative wave--numbers means that the reflection is zero for incidence from the left of all the profiles captured in this region, and this is independent of the value of \(k_{y}\) (because \(k_{y}\) doesn't change the position of the poles in (\ref{helmholtz})). \emph{This means that all permittivity profiles that are analytic at complex positions above the propagation axis do not reflect radiation incident from the left}.  Figures~\ref{branch_cut_fig}i and~\ref{branch_cut_fig}ii confirm this fact for the simple case of a single pole in the permittivity profile.  Figure~\ref{branch_cut_fig}i illustrates wave propagation above the pole at \(z_{0}\) where the reflection is zero, and figure~\ref{branch_cut_fig}ii shows propagation below the pole, where the permittivity varies less rapidly but strong reflection is evident.  Appendix~\ref{appA} contains an analytic demonstration of this phenomenon in a simple exactly solvable case.
%
%
	\section{No branch cuts, no reflection\label{nbnr}}
	\par
	Given that the branch cuts in \(\varphi\) are intimately connected to the phenomenon of reflection, we now investigate under what circumstances they disappear and thereby determine a large set of inhomogeneous media from which the reflection is zero (remembering that the complex position plane describes wave propagation in a family of inhomogeneous media).  To avoid confusion we must emphasise that while the disappearance of the branch cuts is sufficient to remove reflections, it is not necessary: we know that if there is reflection then there will be a branch cut, but this doesn't imply that if there is a branch cut then there is reflection.  We assume an ansatz for the wave \(\varphi\) that is free of branch cuts and therefore by definition free from reflection,
	\begin{equation}
		\varphi(z)=F(z){\rm e}^{{\rm i}k z}\label{ans}
	\end{equation}
	The function \(F(z)\) is assumed to be without branch cuts, with zeros at isolated points \(r_{i}\) and poles at points \(s_{i}\).  Assuming that \(F\to1\) as \(|z|\to\infty\), consistent with the form of \(\epsilon(z)\) given in (\ref{eps_gen}), we write
	\begin{equation}
		F(z)=\prod_{i=0}^{M}\frac{(z-r_{i})^{m_{i}}}{(z-s_{i})^{n_{i}}}.\label{F_fun}
	\end{equation}
	where \(m_{l}\) and \(n_{l}\) are integers such that the numerator and denominator of (\ref{F_fun}) are polynomials of the same degree, and \(M\) is an integer which can be formally taken to infinity.  Inserting (\ref{ans}) in (\ref{helmholtz}) with \(\mu=1\), we find that for such a form of \(\varphi\) the permittivity must equal
	\begin{align}
		\epsilon(z)&=1-\frac{1}{k_{0}^{2}}\left[\frac{F''(z)}{F(z)}+2{\rm i}k\frac{F'(z)}{F(z)}\right]\nonumber\\
		&=1-\frac{1}{k_{0}^{2}}\left[\frac{dG(z)}{dz}+[G(z)]^{2}+2{\rm i}kG(z)\right]\label{eps-G-exp}
	\end{align}
	where \(F'=GF\), with
	\begin{equation}
		G(z)=\sum_{l}\left[\frac{m_{l}}{z-r_{l}}-\frac{n_{l}}{z-s_{l}}\right]\label{G-exp}.
	\end{equation}
	Therefore, after constructing a function \(G(z)\) as the series of simple poles (\ref{G-exp}) with \(\sum m_{l}=\sum n_{l}\), the permittivity profile (\ref{eps-G-exp}) will be such that \(\varphi\) is free from branch cuts and the reflection vanishes for the whole complex position plane.  However, in general this is dependent on the value of \(k_{y}\) because (\ref{eps-G-exp}) explicitly depends on \(k\).  We now show how this angle dependence can be eliminated to give zero reflection for all angles of incidence.  Separating out the expression into a sum of first and second order poles, the expression for the permittivity becomes
	\begin{equation}
		\epsilon(z)=1-\frac{1}{k_{0}^{2}}\sum_{l}\left[\frac{m_{l}(m_{l}-1)}{(z-r_{l})^{2}}+\frac{n_{l}(n_{l}+1)}{(z-s_{l})^{2}}+\frac{a_{l}}{z-r_{l}}+\frac{b_{l}}{z-s_{l}}\right]\label{eps_profile}
	\end{equation}
	where
	\begin{align}
		a_{l}&=2 m_{l}\left[{\rm i}k+\sum_{p\neq l}\frac{m_{p}}{r_{l}-r_{p}}-\sum_{p}\frac{n_{p}}{r_{l}-s_{p}}\right]\nonumber\\
		b_{l}&=2 n_{l}\left[-{\rm i}k+\sum_{p\neq l}\frac{n_{p}}{s_{l}-s_{p}}+\sum_{p}\frac{m_{p}}{r_{p}-s_{l}}\right]\label{coeffs}
	\end{align}
	\begin{figure}[hc!]
		\includegraphics[width=17cm]{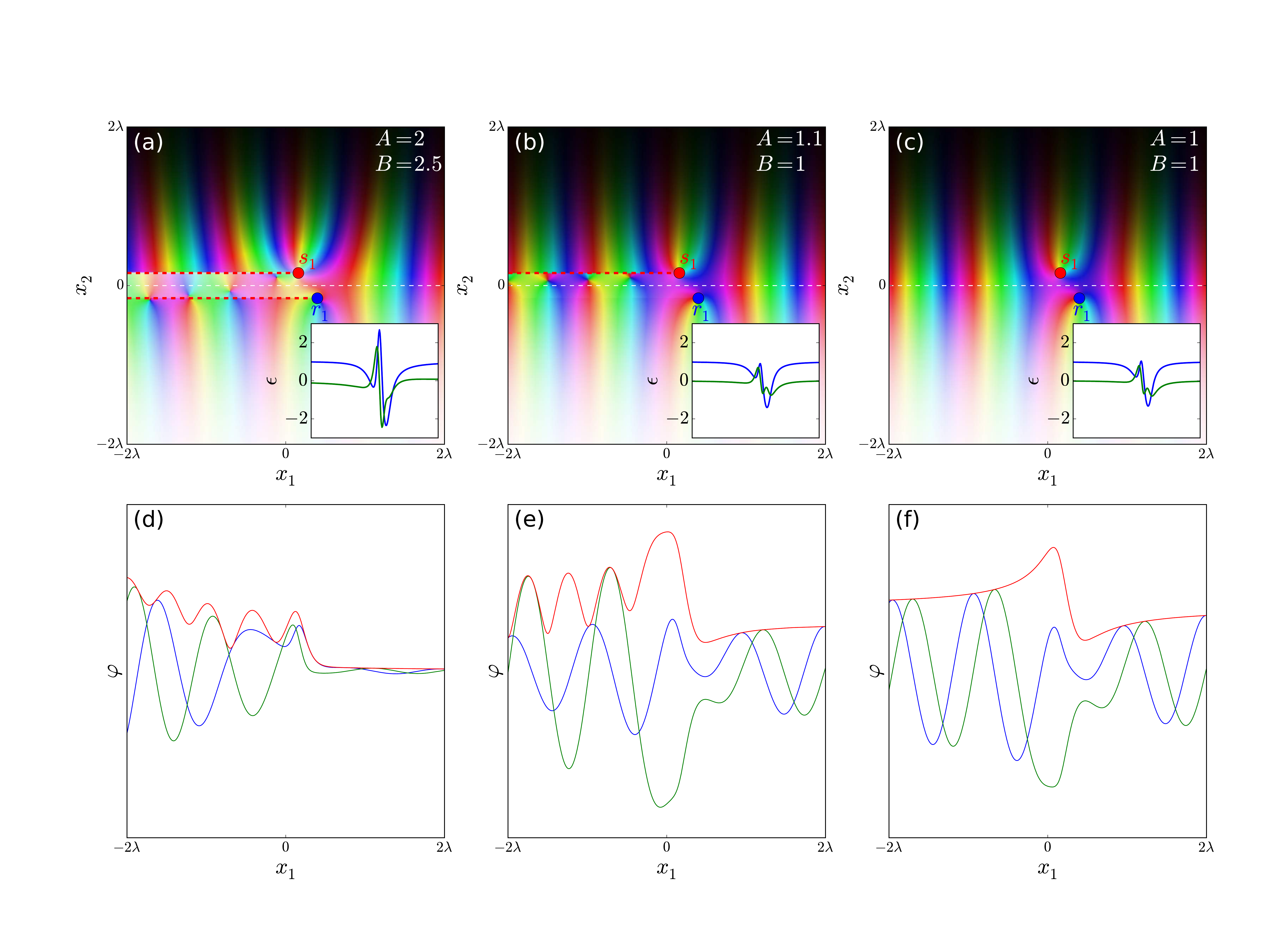}
		\caption{Complex plane representation of wave propagation in the permittivity profile \(\epsilon(z)=1-\frac{2}{k_{0}^{2}}\left[\frac{B}{(z-s_{1})^{2}}+a\left(\frac{1}{z-r_{1}}-\frac{A}{z-s_{1}}\right)\right]\) where \(a=[{\rm i}k(r_{1}-s_{1})-1]/(r_{1}-s_{1})\), with \(r_1=5/2-\ii\) and \(s_1=1+\ii\).  This permittivity reduces to (\ref{eps-profile-11}) when \(A=B=1\).  Panels (a-c) show that as \(A\) and \(B\) are both brought towards unity the branch cuts emanating from the poles in the permittivity at \(r_{1}\) and \(s_{1}\) disappear.  The insets of each of these panels show the permittivity evaluated along the real line.  Panels (d-f) show the wave propagation evaluated along the real axis in each of these cases, and they demonstrate that once the branch cuts have disappeared (strictly speaking only the branch cut from \(s_{1}\) has to disappear), the reflection from the profile vanishes.  Note that in this particular case only a very slight change to the form of the permittivity profile between (b) and (c) leads to a large change in the reflection in panels (e) and (f).\label{branch-cut-figure}}
	\end{figure}
	This tells us how to relate the residues of the poles in the permittivity profile to eliminate the branch cuts discussed in the previous section, which is numerically demonstrated in figure~\ref{branch-cut-figure}.  As already discussed, in general (\ref{eps_profile}) is a function of \(k_{y}\), because it depends on \(k\) through \(a_{l}\) and \(b_{l}\).  This means that a different inhomogeneous medium is required to suppress reflection for each angle of incidence.  To eliminate reflection for all angles of incidence requires this \(k_{y}\) dependence to disappear from (\ref{coeffs}), which for example would require \(m_{l}=1\) and a choice of \(r_{l}\) such that \(a_{l}=0\), and that \(b_{l}\) is independent of \(k\).  
	\par
	The simplest example of (\ref{eps_profile}) is to take a single pole and zero in (\ref{F_fun}) with \(m_{1}=n_{1}=1\).  This leads to the following family of reflectionless profiles
	\begin{equation}
		\epsilon(z)=1-\frac{2}{k_{0}^{2}}\left[\frac{1}{(z-s_{1})^{2}}+\frac{{\rm i}k(r_{1}-s_{1})-1}{(r_{1}-s_{1})}\left(\frac{1}{z-r_{1}}-\frac{1}{z-s_{1}}\right)\right]\label{eps-profile-11}
	\end{equation}
	which are complex functions of position along the lines of constant \(x_{2}\) (see figure~\ref{branch-cut-figure}).  In the particular case where \(r_{1}=s_{1}-{\rm i}/k\) we obtain the permittivity profile \(\epsilon(z)=1-2k_{0}^{-2}(z-s_{1})^{-2}\) which is reflectionless for all \(k_{y}\).  Similarly, taking \(n_{1}=n\), \(n_{i>1}=0\), \(m_{i\leq n}=1\) and \(m_{i>n}=0\) we find the permittivity profile
	\[
		\epsilon(z)=1-\frac{1}{k_{0}^{2}}\left[\frac{n(n+1)}{(z-s_{1})^{2}}+\sum_{l}\frac{a_{l}}{z-r_{l}}+\frac{b_{1}}{z-s_{1}}\right]
	\]
	where \(a_{l}=2[{\rm i}k+\sum_{p\neq l}1/(r_{l}-r_{p}) - n/(r_{l}-s_{1})]\) and \(b_{1}=-\sum_{l}a_{l}\).  The \(n\) quantities \(r_{l}\) can be chosen so that \(a_{l}=0\) for all \(l\), and then \(b_{1}\) is automatically zero.  One therefore finds that \(\epsilon(z)=1-n(n+1)k_{0}^{-2}(z-s_{1})^{-2}\) is reflectionless for all angles of incidence when \(n\) is an integer.  These omni--directional reflectionless profiles were investigated some time ago by Berry and Howls~\cite{berry1990} in their considerations of the WKB approximation applied to the P\"oschl--Teller potential~\cite{poschl1933}, and they are also a special case of the functions discussed in appendix~\ref{appA}.
	\par
	The aforementioned reflectionless complex profile is actually a constitutive element of the P\"oschl--Teller profile~\footnote{The square of the hyperbolic secant can be represented as the infinite sum \(\sec^{2}(x)=\sum_{k=-\infty}^{\infty}\frac{1}{((k-\frac{1}{2})\pi-x)^{2}}\)~\cite{gradshteyn2000} (JO (451)a).}, which is also a special case of (\ref{eps_profile}--\ref{coeffs}).  Taking \(m_{i}=1\), \(s_{l n}=(l+1/2){\rm i}\pi a\) and \(n_{l n}=n\) (all other \(s_{i}\) and \(n_{i}\) are zero) we have
	\[
		\epsilon(z)=1+\frac{n(n+1)}{k_{0}^{2}a^{2}{\rm cosh}^{2}(z/a)}-\frac{1}{k_{0}^{2}}\sum_{l=-\infty}^{\infty}\left[\sum_{p=0}^{n-1}\frac{a_{n l+p}}{z-r_{n l+p}}+\frac{b_{nl}}{z-{\rm i}(l+\frac{1}{2})\pi a}\right]
	\]
	Assuming \(r_{l n+p}=(l+1/2){\rm i}\pi a+\alpha_{p}\), the coefficients \(a_{l}\) and \(b_{l}\) become
	\begin{align}
		a_{ln+p}&=2\left[{\rm i}k-\frac{1}{a}\sum_{\substack{u=0\\u\neq p}}^{n-1}{\rm coth}\left(\frac{\alpha_{u}-\alpha_{p}}{a}\right)-\frac{n}{a}{\rm coth}\left(\frac{\alpha_{p}}{a}\right)\right]\nonumber\\
		b_{l n}&=2 n\left[-{\rm i}k+\frac{1}{a}\sum_{u=0}^{n-1}{\rm coth}\left(\frac{\alpha_{u}}{a}\right)\right]=-\sum_{p=0}^{n-1}a_{ln + p}
	\end{align}
	In the same way as previously discussed, the \(\alpha_{i}\) can be chosen to make the \(a_{i}\) zero.  Having done this the \(b_{i}\) are also automatically zero.   Thus from the assumption of the absence of branch cuts in the complex position plane we reproduce the result that the P\"oschl--Teller profile \(\epsilon(x)=1+n(n+1)(k_{0}a)^{-2}\cosh^{2}(x/a)\) is reflectionless for all \(k_{y}\).  Indeed, the above analysis does not rely on any assumptions on the value of \(a\), nor on the choice of origin of the complex position plane.  Therefore if we choose \(a=|a|\exp({\rm i}\theta)\) as any complex number, and replace \(z\) by \(z-z_{0}\), where \(z_{0}=|z_{0}|\exp({\rm i}\phi)\), then the P\"{o}schl--Teller potential remains reflectionless for all angles of incidence
	\begin{equation}
		\epsilon(z)=1+\frac{e^{-2{\rm i}\theta}n(n+1)}{k_{0}^{2}|a|^{2}\cosh^{2}[(z-|z_{0}|e^{{\rm i}\phi})e^{-{\rm i}\theta}/|a|]}\label{complex-pt}
	\end{equation}
	a fact that cannot be established on the basis of the inverse scattering method~\cite{wu2011,dodd1982}, because it assumes real values for the permittivity.
	\begin{figure}[hc!]
		\includegraphics[width=16.5cm]{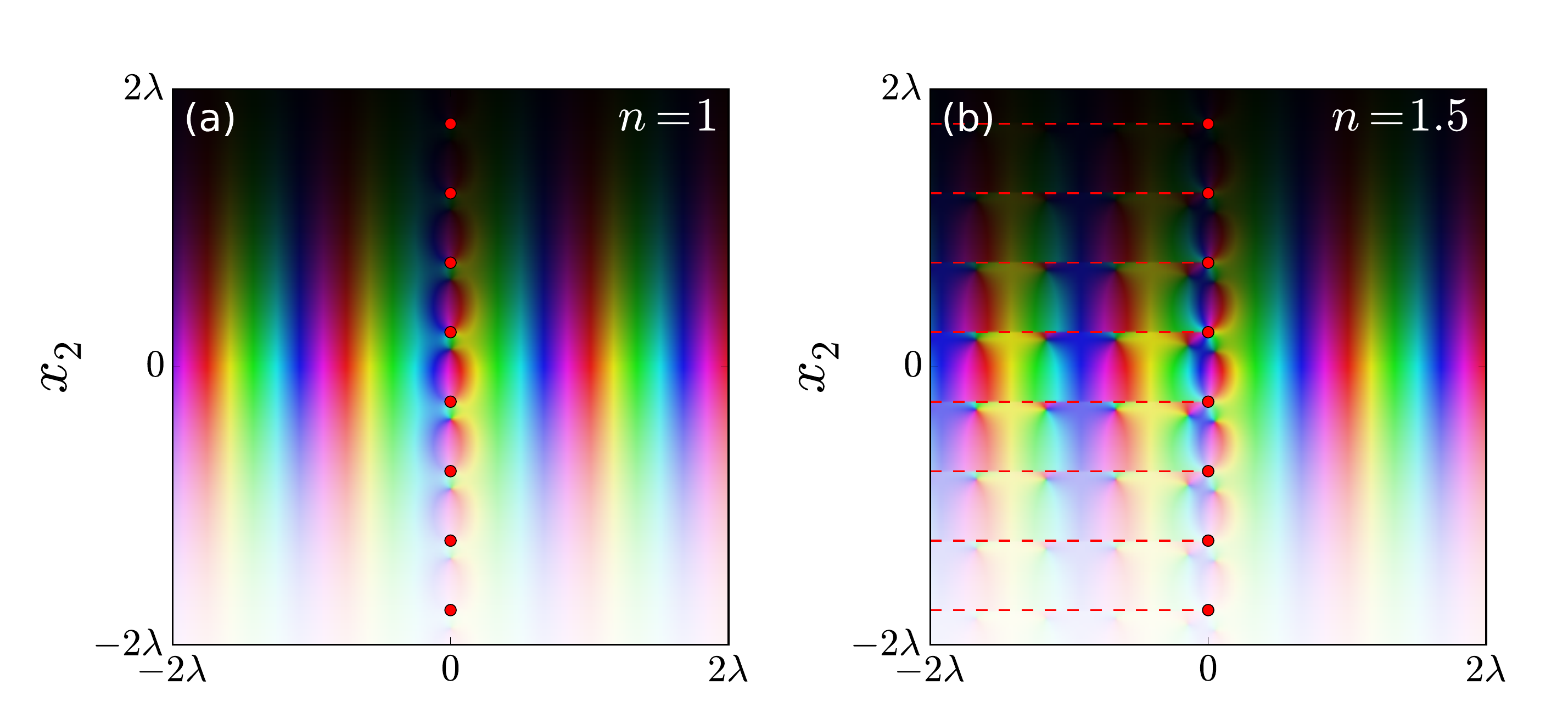}
		\caption{The P\"oschl--Teller potential (\ref{complex-pt}) is reflectionless for all angles when \(n\) is an integer.  The profile is made up of a line of poles parallel to the imaginary position axis.  (a) shows wave propagation in this potential when \(n=1\), demonstrating the lack of any branch cuts in the wave when the profile is reflectionless.  Meanwhile (b) shows wave propagation when \(n\) is non--integer, showing that branch cuts emerge from each of the poles.}
	\end{figure}
%
%
	\section{Relation to the WKB method}
	\par
	The phase integral (WKB) method~\cite{heading2013} already uses an analytic continuation of the wave equation into complex coordinates.  At first sight the findings of this paper seem at odds with the known results of this method, which emphasize that the zeros of \(\epsilon(z)\) rather than the poles are most important for determining the reflection.  In this section we show how to interpret our findings in terms of the WKB solutions to the wave equation.
	\par
	For now, consider the case where \(k_{y}=0\). The WKB approximations to the solutions of the wave equation (\ref{helmholtz}) may be found in standard texts on phase integral methods, e.g.~\cite{heading2013} and are
    \begin{equation}
        \varphi_{\pm}(z)=\frac{1}{\epsilon(z)^{\frac{1}{4}}}\e^{\pm\ii k_{0}\int_{z_{r}}^{z}\sqrt{\epsilon\left(z'\right)}dz'}.\label{WKB}
    \end{equation}
	These expressions are valid approximations when \(k_{0}\gg \epsilon'/\epsilon^{3/2}\); i.e. when the permittivity varies significantly on a scale much larger than the wavelength (in particular, these expressions are an excellent approximation for large \(|z|\), where \(\epsilon\) varies slowly and is close to unity).  The reference point \(z_{r}\) is where the phase of the wave is zero, and we choose this to be a complex position at which \(\epsilon(z_{r})=0\). For the sake of consistency with the literature~\cite{heading2013} we will denote the two expressions in (\ref{WKB}) by \(\varphi_{+}=(z_{r},z)\) and \(\varphi_{-}=(z,z_{r})\).
\par
\begin{figure}[h!]
        \begin{center}
	\includegraphics[width=16cm]{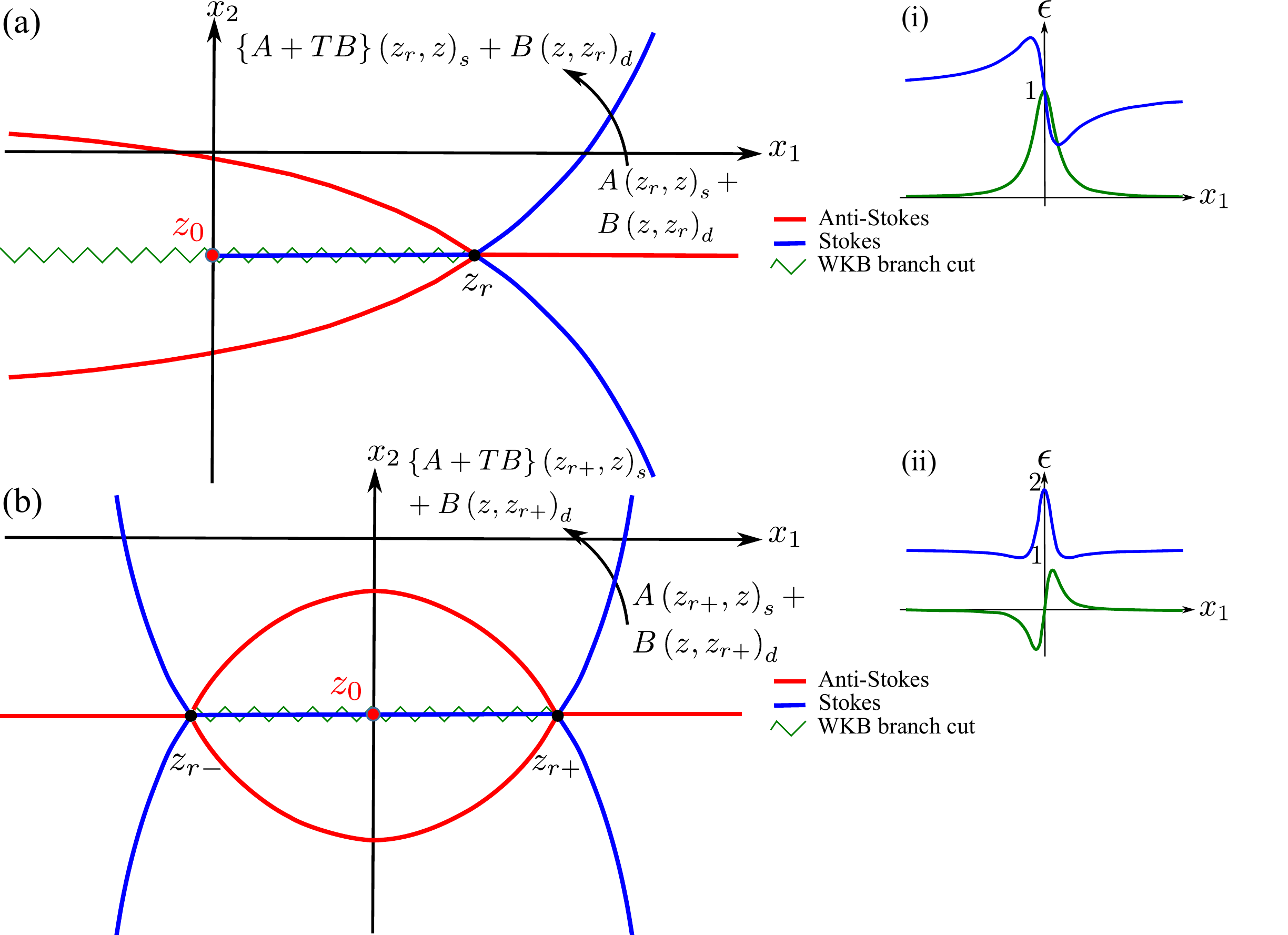}
        \caption{The Stokes and anti-Stokes lines for the two permittivity profiles (a) \(\epsilon(z)=1-\frac{1}{z+\ii}\) and (b) \(\epsilon(z)=1-\frac{1}{\left(z+\ii\right)^{2}}\) (subscripts `s' and `d' indicate the relative dominancy of the two WKB solutions, and insets (i) and (ii) show the permittivity profiles along the real line, with the same colour conventions as in the previous figures). Both panels illustrate change in a general WKB approximation taken from the right anti-clockwise about \(z_{r}\) across the Stokes line.  Both profiles show the same behaviour illustrated in figure~\ref{noreflection}.  The small amplitude right propagating wave is unchanged across this curve, whereas the large amplitude left propagating wave undergoes a change which depends on a constant \(T\), called the Stokes constant associated with the line.  For completeness we have included the branch cuts that occur in the WKB approximations (\ref{WKB}), which are purely due to the approximate nature of these expressions~\cite{heading2013}.\label{pole}}
        \end{center}
    \end{figure}

In the WKB method, the phenomenon of reflection is associated with the breakdown of the approximation at the zeros of \(\epsilon(z)\), and the complex positions of these zeros are often used to calculate reflection coefficients~\cite{landau2003}. In order to find the correct WKB approximation to a particular exact solution to the wave equation, we must use a patchwork of different linear combinations of the two WKB approximations \(\varphi=A(z_{r},z)+B(z,z_{r})\) throughout the complex position plane.  Ultimately it is the changes in these coefficients \(A\) and \(B\) as we move through the complex position plane that determines the amount of reflection from the medium.  The change in these coefficients is known as the \emph{Stokes phenomenon}, and occurs across what are known as \emph{Stokes} lines, defined as the curves in the complex plane satisfying
    \begin{equation}
        \text{Re}\left(\int_{z_{r}}^{z}\sqrt{\epsilon\left(z'\right)}dz'\right)=0\label{Stokes}
    \end{equation}
    complementary to these lines are the \emph{anti--Stokes} lines, defined as the curves satisfying
        \begin{equation}
    \begin{split}
        \text{Im}\left(\int_{z_{r}}^{z}\sqrt{\epsilon\left(z'\right)}dz'\right)=0\label{anti-Stokes}.
    \end{split}
    \end{equation}
	On the anti-Stokes lines both \((z_{r},z)\) and \((z,z_{r})\) have the same magnitude, because the phase is purely real.  These curves divide the complex plane into regions where one of the two WKB solutions has a larger amplitude than the other.  Typically the larger of the two is called the \emph{dominant} solution, and the smaller is called the \emph{sub--dominant} solution.  The Stokes lines are the curves along which the amplitude of the dominant WKB solution is maximal compared to the sub--dominant solution.  In general as we cross a Stokes line the error inherent in the WKB expression for the dominant wave will exceed the magnitude of the subdominant wave.  In this region the coefficient of the subdominant wave is undetermined, and in general it must be changed after one has passed through the Stokes line in order that the patchwork of WKB solutions represent a good approximation to the exact solution.  In WKB theory, reflection is associated with this change in the coefficient of the subdominant wave across a Stokes line, as illustrated in figure~\ref{pole}.
This is the Stokes phenomenon. As \(|z|\to\infty\), \(\epsilon(z)\to 1\) and we can see from the definitions (\ref{Stokes}--\ref{anti-Stokes}) that the Stokes and anti-Stokes lines will asymptote to lines parallel to the imaginary and real position axis, respectively.
\par
\begin{figure}[h!]
        \begin{center}
	\includegraphics[width=16cm]{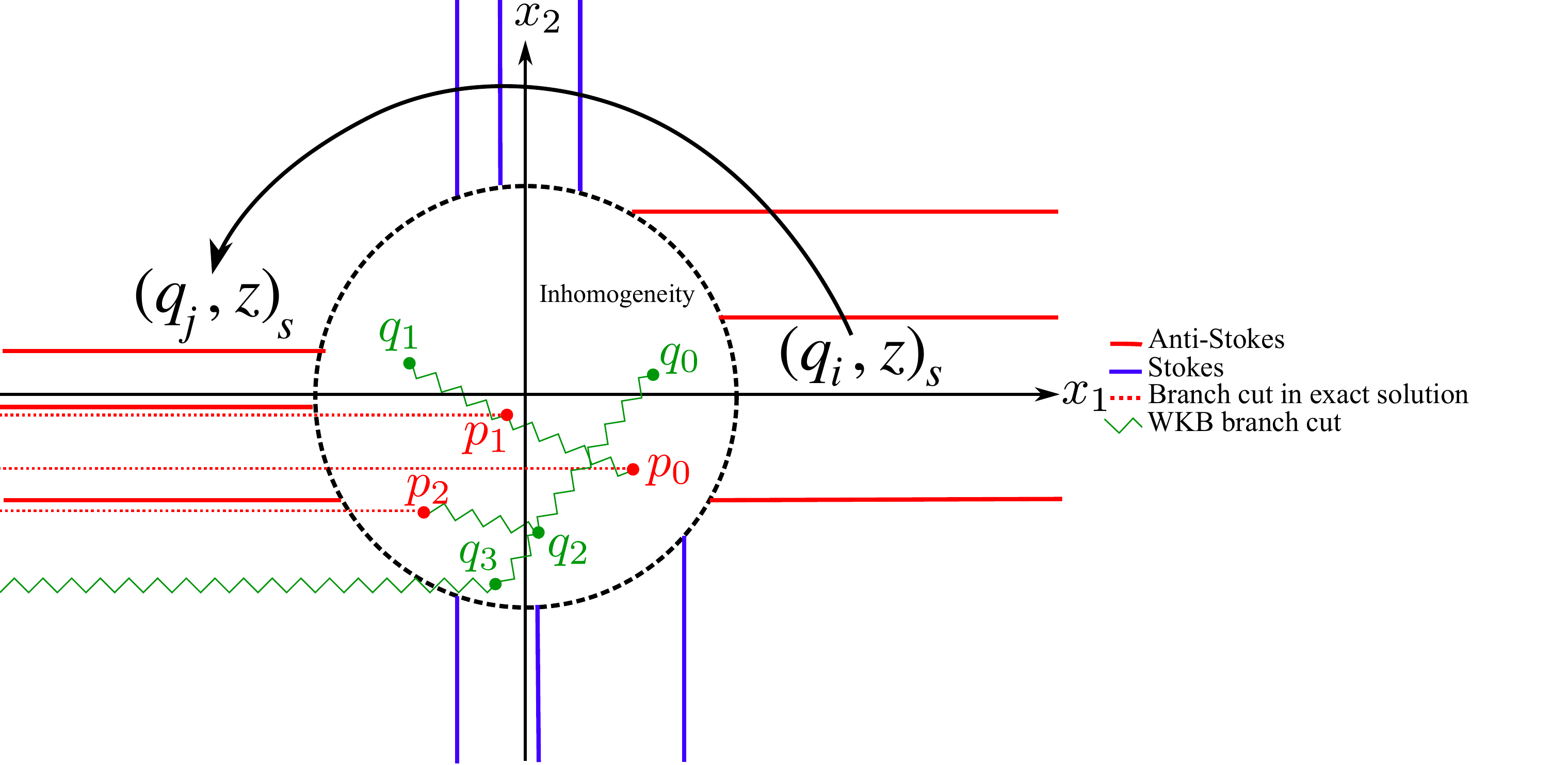}
        \caption{A schematic diagram of the complex position plane for a profile analytic in the upper half plane, containing poles at \(p_{i}\) and zeros at \(q_{j}\). Outside the dashed black circle, the permittivity is approximately one, so the Stokes and anti-Stokes lines are approximately horizontal and vertical, respectively. When the radius of the semi-circle is sufficiently large, we will cross a number of anti-Stokes lines, followed by a number of Stokes lines and then more anti-Stokes lines before reaching the negative real axis. Just before encountering the Stokes lines, \((z_{r},z)\) will be a subdominant (low amplitude wave). Because of this subdominancy, crossing the Stokes lines will not change the wave and we arrive at the negative real position axis with the solution remaining solely as a right travelling wave.  Hence we get zero reflection, in agreement with the argument of section~\ref{pole-section}.\label{noreflection}}
        \end{center}
    \end{figure}

To compare with the findings of the previous sections, we now consider a permittivity profile containing poles only in the lower half position plane, which we have suggested ought to be reflectionless for all angles of incidence. In figure~\ref{noreflection} we consider a purely right travelling wave on the far right (\(x_{1}\to\infty\)) of a generic profile containing poles in the lower half position plane.  Moving along a large semi-circle in the upper half plane (where a right travelling wave becomes subdominant before encountering any Stokes lines), we find that the configuration of Stokes and anti--Stokes lines as \(|z|\to\infty\) guarantees zero reflection~\footnote{There are a few subtleties that should be explained before accepting the simplicity of this result. Firstly, the change in the subdominant coefficient occurs across a Stokes line of the particular reference point (a zero) of the permittivity. In general, the dominancy of the WKB approximations may alter when we integrate between reference points, which could introduce a reflected wave into our approximation. However, this does not happen in this situation because, having taken the radius of the semi-circle to be very large, the right going WKB solution will be subdominant with respect to any reference point which is a zero of \(\epsilon(z)\). Secondly, the WKB approximations have branch cuts (separate from the branch cuts of the exact solution). In general all zeros and poles of \(\epsilon(z)\) and \(z=\infty\) will be branch points of (\ref{WKB}). However, we can always ensure that we do not cross any of these branch cuts in tracing around the semi-circle by placing the branch cut to \(z=\infty\) in the lower half position plane, as shown in figure~\ref{noreflection}.}.  Notice that the branch cuts emerging from the poles that we discussed previously must be incorporated into the WKB theory if we are to avoid the conclusion that the medium is also reflectionless for waves incident from the right (see~\cite{horsley2015}).
\par
 The WKB method is therefore consistent with the previous sections, where we found that the region above the poles in \(\epsilon(z)\) corresponds to a family of reflectionless inhomogeneous media for radiation incident from the left.  Moreover, via this method we can also gain further information about the transmission through these profiles. When considering a complex valued permittivity the position of the reference point \(z_{r}\) in (\ref{WKB}) is not arbitrary.  This is because both amplitude and phase change with \(z_{r}\).  We take the reference point such that a right--going wave has unit amplitude on the far left of the profile \(x_{1}\to-\infty\) (shown as point \(b\) in figure~\ref{noreflection}).  The amplitude of the wave on the far right of the profile (point \(a\) of figure~\ref{noreflection}) then gives us the transmission coefficient,
    \begin{equation}
        t=\lim_{\substack{a\to\infty\\b\to-\infty}}\e^{\ii k_{0}\int_{b}^{a}\sqrt{\epsilon(z)}dz}.\label{transmission1}
    \end{equation}
Note that the argument of this quantity will not converge, so we cannot strictly give a meaning to the phase of the transmission coefficient through such infinitely extended profiles. However, the time average of the transmitted power is meaningful, which is
    \begin{equation}
        |t|^{2}=\e^{-2k_{0}\text{Im}\left(\int_{C}\sqrt{\epsilon(z)}dz\right)}.\label{transmission2}
    \end{equation}
where the integration contour has been deformed into a clockwise semi--circle in the upper half position plane.  Note that for a permittivity profile composed of a collection of poles in the lower half plane, \(\epsilon(z)=1+\alpha_{1}/(z-z_{1})+\dots+\alpha_{n}/(z-z_{n})+\beta_{1}/(z-z_{n+1})^{2}+\dots+\beta_{n}/(z-z_{n+m})^{2}+\dots\), (\ref{transmission2}) depends only on the sum of the residues of the simple poles
\[
	|t|^{2}=\e^{k_{0}\pi\sum_{i}\alpha_{i}}
\]
Therefore, for a permittivity profile that is analytic above the line of propagation, in addition to zero reflection for incidence from the left, the transmission is unity when the sum of the residues of the simple poles of \(\epsilon(z)\) is zero.  The reflection coefficients are less trivial to find because they are dependent on the values of the Stokes constants, which in general we do not know.
%
%
	\section{Summary and conclusions}
	\par
	In this work we have explored the utility of analytically continuing wave equations into the complex position plane.  In a similar spirit to the formalism of transformation optics, we found that in electromagnetism there is a simple interpretation for the entire complex position plane, where each curve \(z(\gamma)\) can be understood as a different inhomogeneous anisotropic medium.  We then investigated the properties of solutions to the one dimensional Helmholtz equation \(\varphi''+[k_{0}^{2}\epsilon(z)-k_{y}^{2}]\varphi=0\) in the whole complex position plane, using a combination of exact and approximate analytical techniques, including the WKB method.
	\par
	Considering bounded permittivity profiles constructed as a sum of poles of varying degrees, we found that in general there are branch cuts in the wave \(\varphi(z)\) that emanate from the poles of \(\epsilon(z)\).  These branch cuts are connected to the reflection of the wave.  In the region of the complex position plane that lies above the poles in the permittivity there is no reflection, whereas in general below there is, and the branch cuts account for this jump.  We found that this knowledge can be used to construct reflectionless permittivity profiles in two distinct ways.  We can either construct a profile where all the poles occur in the region of complex position below the axis of propagation (which reproduces the recent findings of~\cite{horsley2015}), or we can demand that the branch cuts disappear, which we have shown reproduces a generalisation of the P\"oschl--Teller potential, equivalent to a permittivity profile in optics.
	\acknowledgments
    	The authors acknowledge useful discussions with I. R. Hooper, J. R. Sambles and W. L. Barnes.  SARH and TGP acknowledge financial support from EPSRC program grant EP/I034548/1.
%
%
	\appendix
	\section{An analytic solution\label{appA}}
	\par
	In this appendix we confirm---using a particular family of exact solutions to (\ref{helmholtz})---that branch cuts in the wave emerge from poles in the permittivity.  The particular equation of interest is a type of confluent hypergeometric equation: Whittaker's differential equation~\cite{dlmf2015}, which takes the form
	\begin{equation}
		\frac{d^{2}\varphi}{d\zeta^{2}}+\left(-\frac{1}{4}+\frac{\kappa}{\zeta}+\frac{\frac{1}{4}-\mu^{2}}{\zeta^{2}}\right)\varphi=0\label{whit_eqn}
	\end{equation}
	and has two standard solutions, \(W_{\pm\kappa,\mu}(\pm \zeta)\).  Changing variables \(\zeta=2{\rm i}k_{0}(z+{\rm i}z_{0})\) we find a differential equation in the form of (\ref{helmholtz}) with \(\mu=1\), \(k_{y}=0\) and a permittivity
	\[
		\epsilon(z)=1+\frac{2{\rm i}\kappa}{k_{0}(z+{\rm i}z_{0})}+\frac{\frac{1}{4}-\mu^{2}}{k_{0}^{2}(z+{\rm i}z_{0})^{2}}.
	\]
	which has a pole at the complex position \(-{\rm i}z_{0}\).  Close to \(-\ii z_{0}\) the solutions take a particularly simple form where the branch cut can be seen immediately.  When \(\mu\neq1/2\), the double pole dominates and the solution approximates to
	\begin{equation}
		\varphi\sim(z-{\rm i}z_{0})^{\pm\mu + \frac{1}{2}}\label{approx-1}
	\end{equation}
	and when \(\mu=1/2\), the simple pole dominates and the solution is
	\begin{equation}
		\varphi\sim\begin{cases}
			\frac{{\rm i}}{2k_{0}\kappa}+(z-{\rm i}z_{0})\ln\left(k_{0}(z-{\rm i}z_{0})\right)\\
			\left(z-{\rm i}z_{0}\right)-{\rm i}\kappa k_{0}\left(z-{\rm i}z_{0}\right)^{2}
			\end{cases}\label{approx-2}
	\end{equation}
	Notice that in general both (\ref{approx-1}) and (\ref{approx-2}) have branch cuts emerging from \(-\ii z_{0}\).  These are the branch cuts discussed in the main text.  Now we examine the opposite case, and find the behaviour of these branch cuts far from the pole.
	\par
	The exact solutions \(W_{\pm\kappa,\mu}(\pm \zeta)(z)\) are cut along the line \({\rm arg}(z)=\pm\pi\), and the general formula to continue the wave across this cut onto the next Riemann sheet is given in terms of the Whittaker functions evaluated on the first sheet by~\cite{dlmf2015}
	\begin{multline}
		(-1)^{m}W_{\kappa,\mu}(z{\rm e}^{2m\pi\ii})=-\frac{{\rm e}^{2\kappa\pi\ii}\sin(2m\mu\pi)+\sin((2m-2)\mu\pi)}{\sin(2\mu\pi)}W_{\kappa,\mu}(z)\\
		-\frac{\sin(2m\mu\pi)2\pi\ii {\rm e}^{\kappa\pi\ii}}{\sin(2\mu\pi)\Gamma\left(\frac{1}{2}+\mu-\kappa\right)\Gamma\left(\frac{1}{2}-\mu-\kappa\right)}W_{-\kappa,\mu}(z{\rm e}^{\pi\ii})\label{analytic-continuation}
	\end{multline}
	Now consider the form of these functions along the real \(z\) axis at \(\pm\infty\).  To find this we use the asymptotic forms of the Whittaker functions on the first sheet
	\begin{equation}
		W_{\kappa,\mu}(|z|\to\infty)\sim{\rm e}^{-\frac{1}{2}z}z^{\kappa}\label{asymp_exp}.
	\end{equation}
	We take the particular case of \(\mu=\frac{1}{2}\) and \(\kappa=\frac{\ii k_{0}A}{2}\).  For the solution \(W_{-\frac{\ii k_{0}A}{2},\frac{1}{2}}(-2\ii k_{0}(z+\ii z_{0}))\), with \(z_{0}>0\) the argument remains on the first sheet as we trace the solution from \(z=+\infty\) to \(z=-\infty\).  Therefore asymptotically the wave is
	\begin{equation}
		W_{-\frac{\ii k_{0}A}{2},\frac{1}{2}}(-2{\rm i}k_{0}(z+{\rm i}z_{0}))\sim\begin{cases}
		{\rm e}^{\ii k_{0}(z+\ii z_{0})}{\rm e}^{-\frac{\ii k_{0}A}{2}[\log(2 k_{0}|z+\ii z_{0}|)-\frac{\ii\pi}{2}]}&z\to+\infty\\
		{\rm e}^{\frac{k_{0}\pi A}{2}}{\rm e}^{\ii k_{0}(z+\ii z_{0})}{\rm e}^{-\frac{\ii k_{0}A}{2}[\log(2 k_{0}|z+\ii z_{0}|)-\frac{\ii\pi}{2}]}&z\to-\infty
		\end{cases}\label{above_pole}
	\end{equation}
	which is right--going on both sides and exhibits no reflection.  Meanwhile if \(z_{0}<0\), we pass through the branch cut as we follow the solution back to \(z\to-\infty\).  Applying (\ref{analytic-continuation}) we can move the cut parallel to the real \(z\) axis and we find the asymptotic forms
	\begin{equation}
		W_{-\frac{\ii k_{0}A}{2},\frac{1}{2}}(-2{\rm i}k_{0}(z+{\rm i}z_{0}))\sim
		{\rm e}^{\ii k_{0}(z+\ii z_{0})}{\rm e}^{-\frac{\ii k_{0}A}{2}[\log(2 k_{0}|z+\ii z_{0}|)-\frac{\ii\pi}{2}]}\qquad z\to+\infty\label{below_pole1}
	\end{equation}
	and
	\begin{multline}
		W_{-\frac{\ii k_{0}A}{2},\frac{1}{2}}(-2{\rm i}k_{0}(z+{\rm i}z_{0}))\sim
		{\rm e}^{-\frac{k_{0}A\pi}{2}}{\rm e}^{\ii k_{0}(z+\ii z_{0})}{\rm e}^{-\frac{\ii k_{0}A}{2}[\log(2k_{0}|z+\ii z_{0}|)-\frac{\ii\pi}{2}]}\\
		-\frac{2\pi\ii}{\Gamma(\frac{\ii k_{0}A}{2})\Gamma(1+\frac{\ii k_{0}A}{2})}{\rm e}^{-\ii k_{0}(z+\ii z_{0})}{\rm e}^{\frac{\ii k_{0}A}{2}[\log(2k_{0}|z+\ii z_{0}|)+\frac{\ii \pi}{2}]}\qquad z\to-\infty\label{below_pole2}
	\end{multline}
	As stated in the main text, the presence of the cut is clearly connected to the phenomenon of reflection.  As we move from above to below the pole at \(-\ii z_{0}\), the solution to (\ref{whit_eqn}) goes from (\ref{above_pole}) to (\ref{below_pole1}--\ref{below_pole2}), and on the left we now have a reflected wave.  The above analysis enables to identify the reflection and transmission coefficients of this extended profile, which are (above the cut)
	\begin{align}
		T_{>}&={\rm e}^{-\frac{k_{0}\pi A}{2}}\nonumber\\
		R_{>}&=0\label{tra}
	\end{align}
	and (below the cut)
	\begin{align}
		T_{<}&={\rm e}^{\frac{k_{0}\pi A}{2}}\nonumber\\
		R_{<}&=\left|\frac{2\pi {\rm e}^{\frac{k_{0}\pi A}{2}}}{\Gamma\left(\frac{\ii k_{0}A}{2}\right)\Gamma\left(1+\frac{\ii k_{0}A}{2}\right)}\right|=2{\rm e}^{\frac{k_{0}\pi A}{2}}\sinh(\frac{\pi k_{0}A}{2})\label{trb}
	\end{align}
	where to obtain the second of (\ref{trb}) we applied the properties of the \(\Gamma\) function given in \cite{dlmf2015} (eq. 5.4.3).  To first order in \(A\) these reflection and transmission coefficients are \(T_{>/<}\sim1\mp\frac{k_{0}\pi A}{2}\) and \(R_{<}\sim\pi k_{0} A\), in agreement with the results of the Born approximation given in (\ref{reflection-complex-plane}--\ref{transmission-complex-plane}).
	
	In cases where \(\kappa=0\) and \(\mu\neq0\) in (\ref{whit_eqn}), the permittivity approaches unity as \(1/z^{2}\) and the reflection coefficient \(R_{<}\) has a denominator containing \(\Gamma(\frac{1}{2}\pm\mu)\).  When \(\mu\) equals an integer plus one half we are at a pole of the \(\Gamma\) function, and the reflection coefficient vanishes.  In this case the branch cut also vanishes, and we reproduce the reflectionless behaviour found in section~\ref{nbnr}.

\end{document}